\documentclass[a4paper,11pt]{article}
\usepackage{atlasphysics}
\usepackage{jheppub} 
\usepackage{hyperref}
\usepackage{placeins} 
\usepackage{subfigure}
\usepackage{slashed}
\usepackage{xspace}
\usepackage{verbatim} 

\usepackage{graphicx} 
\usepackage{mathptmx} 
\usepackage{epsfig}

\usepackage{acronym}
\usepackage{verbatim}
\usepackage{lscape}

\usepackage{float}
\usepackage[countmax]{subfloat}
\usepackage[numbers,sort&compress]{natbib}

\def \signalyield{$27.4^{+7.5} _{-6.5} $}
\def \origsignalyield{$29.2^{+7.5} _{-6.5} $}
\def \significance{$5.1\sigma$}

\def \zmumu {$Z \to \mu^+ \mu^-$\xspace}

\def \jpsi {\ensuremath{\Jpsi}\xspace}
\def \w {\ensuremath{W}\xspace}
\def \wpm {$\Wpm$\xspace}
\def \wjpsi {$\Wpm + \Jpsi$\xspace}
\def \wpjpsi {$\Wpm + \textrm{prompt} ~\Jpsi$\xspace}
\def \ifb {fb$^{-1}$\xspace}

\def \WTM {\ensuremath{m_\textrm{T}(\w)}\xspace}

\def \GeV {GeV\xspace}

\newcommand{\ptjpsi}{\ensuremath{\pt^{\jpsi}}\xspace}
\newcommand{\yjpsi}{\ensuremath{y_{\jpsi}}\xspace}

\newcommand{\rfid}{\ensuremath{R_{\jpsi}^\mathrm{fid}}\xspace}
\newcommand{\rincl}{\ensuremath{R_{\jpsi}^\mathrm{incl}}\xspace}
\newcommand{\rdpssub}{\ensuremath{R_{\jpsi}^\textrm{DPS sub}}\xspace}

\usepackage{preprintcover}  
\PreprintCoverPaperTitle{Measurement of the production cross section of prompt \boldmath{\jpsi} mesons 
in association with a \boldmath{\wpm} boson in \boldmath{$pp$}
collisions at \boldmath{$\sqrt{s}=\mathrm{7}$}~TeV with the ATLAS detector}  
\PreprintIdNumber{CERN-PH-EP-2013-184}  
\PreprintCoverAbstract{The process $pp \to \wpm \Jpsi$ provides a powerful probe of the production
mechanism of charmonium in hadronic collisions, and is also sensitive
to multiple parton interactions in the colliding protons. 
Using the 2011 ATLAS dataset of 4.5~\ifb\ of $\sqrt{s} = 7$ TeV $pp$
collisions at the LHC, the
first observation is made of the production of \wpm + prompt
\jpsi\ events in hadronic collisions, using $\wpm \to \mu \nu_{\mu}$ and
$\jpsi \to \mumu$. A yield of \signalyield\ \wpjpsi events is
observed, with a statistical significance of \significance. 
The production rate as a ratio to the inclusive \wpm\ boson production
rate is measured, and the double parton scattering contribution to the cross
section is estimated.
}
\PreprintJournalName{JHEP}  

\begin{document}

%
\title{Measurement of the production cross section of prompt \boldmath{\jpsi} mesons 
in association with a \boldmath{\wpm} boson in \boldmath{$pp$}
collisions at \boldmath{$\sqrt{s}=\mathrm{7}$}~TeV with the ATLAS detector}
%
\author{The ATLAS Collaboration}
\date{March 14, 2014}
\emailAdd{atlas.publications@cern.ch}

\abstract{
The process $pp \to \wpm \Jpsi$ provides a powerful probe of the production
mechanism of charmonium in hadronic collisions, and is also sensitive
to multiple parton interactions in the colliding protons. 
Using the 2011 ATLAS dataset of 4.5~\ifb\ of $\sqrt{s} = 7$ TeV $pp$
collisions at the LHC, the
first observation is made of the production of \wpm + prompt
\jpsi\ events in hadronic collisions, using $\wpm \to \mu \nu_{\mu}$ and
$\jpsi \to \mumu$. A yield of \signalyield\ \wpjpsi events is
observed, with a statistical significance of \significance. 
The production rate as a ratio to the inclusive \wpm\ boson production
rate is measured, and the double parton scattering contribution to the cross
section is estimated.
}

\maketitle
\flushbottom


\section{Introduction}

Study of the production of a $W$ boson in association with a prompt $J/\psi$ meson offers new tests of
Quantum Chromodynamics (QCD) at the perturbative/non-perturbative boundary as well as developing the framework for 
future probes of the Higgs sector and beyond-the-standard-model searches in such final states.

Perturbative calculations of heavy quarkonium production in hadronic collisions distinguish between terms that produce a heavy quark system ($Q\bar{Q}$) in a colour-singlet (CS) or a colour-octet (CO) state.
The relative importance of these terms for inclusive \jpsi\ production is a subject of
debate\,\cite{Brambilla:2004wf,Brambilla:2010cs,Lansberg:2008gk,Butenschoen:2011yh,Butenschoen:2012qr,Baranov:2011ib,Ma:2010jj}.
In the case of prompt \jpsi\ production in association with a \wpm\ boson, the
relative contributions of CS and CO processes differ from the inclusive process.
Some theoretical studies\,\cite{PhysRevD.66.114002,Gang2011} suggest \wpjpsi production should be dominated by colour-octet processes, and thus be a distinctive test
of the non-relativistic QCD (NRQCD) framework\,\cite{Caswell:1985ui,Bodwin:1994jh}. 
In contrast, recent work\,\cite{Lansberg:2013wva} suggests that in 7~TeV $pp$ collisions, 
CO and CS (in particular, electromagnetic $\wpm \gamma^*\to \wpm \Jpsi$) contributions to the \wpjpsi\ cross section are comparable.  
Measurements of the production cross sections can help distinguish between these models.  
A search for the related processes $\wpm+\Upsilon(1S)$ and $\Zzero+\Upsilon(1S)$ performed by the CDF
experiment saw no excess of events above the expected background and set upper limits on the production rate\,\cite{Acosta:2003mu}.

Observation and measurement of \wpjpsi production for the first time represents a step in our understanding toward measurements of the Higgs boson in rare
quarkonia and associated vector boson decay modes, first proposed in Ref.~\cite{Kartvelishvili:1988pu}. Recent phenomenological studies\,\cite{Bodwin:2013gca}
have emphasised the value of these rare decay modes to provide a unique probe of the Higgs boson charm couplings.
Such final states can also be sensitive probes of beyond-the-standard-model (BSM) frameworks.
The presence of an anomalous rate of $\Wpm/Z+\textrm{prompt}~\Jpsi / \Upsilon$ associated production over standard model predictions
can, for example, be an indication of a signature of a charged Higgs boson decay at low $\tan\beta$ in some supersymmetry models\,\cite{Grifols:1987iq},
or explore the possible existence of a new light scalar particle\,\cite{Clarke:2013aya}. 
The complementarity of vector boson plus quarkonia final state measurements to ongoing BSM search programmes
further emphasises the need for a robust understanding of the QCD production modes.

In addition to the single parton scattering (SPS) reaction studied as a probe of quarkonium production, double parton scattering (DPS)
interactions\,\cite{Paver:1982yp,Sjostrand:2004pf,Korotkikh:2004bz,Gaunt:2009re,Bartalini:2011jp},
where the \wpm\ boson and \jpsi are produced in separate parton--parton
collisions from the same proton--proton interaction, can contribute to the total rate
for production of a $\wpm+\jpsi$ final state.  These processes are not
distinguishable on an event-by-event basis from SPS processes, but are
expected to differ in overall kinematic features, such as angular
correlations. Any measurement of the \wjpsi process will include both SPS and DPS contributions, and provides useful information on the DPS process as well.

This paper reports the observation of \wpjpsi production in the
$\wpm\ (\to \mu^{\pm}\nu_{\mu})+\jpsi (\to \mumu)$ channel. 
Prompt \Jpsi\ candidates that are produced in decays of
heavier charmonium states (e.g.\ $\chic \to \gamma + \Jpsi$) are not distinguished from directly-produced \Jpsi.
\wpjpsi events due to DPS are included in the measurement
and their contribution to the total rate is estimated. This production
mechanism (including both SPS and DPS contributions) is separated from the background of $\Wpm + b$-hadrons
with $b \to \Jpsi + X$. An additional background from misidentified
multi-jets is also considered.
The cross-section ratio of \wpjpsi production to the
inclusive \wpm\ production is measured in the fiducial phase space of the \wpm\ boson and \jpsi.
The cross-section ratio is also reported with the \jpsi\ rate corrected for the muons that
fall outside of the detector acceptance in transverse momentum and
pseudorapidity for a given \jpsi\ transverse momentum. This correction depends on the (unknown) spin-alignment of \jpsi\ produced in association with a \wpm\ boson and so the results are provided for the full envelope
of possible spin-alignment scenarios as was previously done for
inclusive \jpsi\ and $\Upsilon$ production measurements\,\cite{Aad:2011sp,Aad2012dlq}.

The data used in this analysis were collected using the ATLAS detector
during the 2011 proton--proton run of the Large Hadron Collider (LHC) at centre-of-mass energy
7~TeV and correspond to an integrated luminosity of $4.51 \pm 0.08$~\ifb \cite{Aad:2013ucp}.

\section{The ATLAS detector}
ATLAS is a multi-purpose detector\,\cite{1748-0221-3-08-S08003},\footnote{ATLAS uses a
  right-handed coordinate system with its origin at the nominal
  interaction point (IP) in the centre of the detector and the $z$-axis
  along the beam pipe. The $x$-axis points from the IP to the centre of
  the LHC ring, and the $y$-axis points upward. Cylindrical coordinates
  ($r$,$\phi$) are used in the transverse plane, $\phi$ being the
  azimuthal angle around the beam pipe, referred to the $x$-axis. The pseudorapidity is defined
  in terms of the polar angle $\theta$ as $\eta =  -\ln\tan(\theta/2)$.}
 designed to study a variety of phenomena at the LHC.
The inner detector (ID), which is surrounded by a superconducting solenoid
that produces a 2~T magnetic field, performs tracking of charged
particles for $|\eta|<2.5$. The calorimeter system covers $|\eta|<4.9$
and detects energy deposits from electrons, photons, muons and
hadrons. The muon spectrometer (MS), outside the calorimeters,
measures muon momenta using air-core superconducting toroidal magnets.
A typical muon traverses three precision position measurement stations
covering $|\eta|<2.7$.  Fast trigger chambers cover $|\eta|<2.4$.  The
MS has a cylindrical barrel geometry covering $|\eta| \lesssim 1$ and
an endcap disk geometry covering $|\eta| \gtrsim 1$.  The combination
of ID and MS tracking reconstructs muons with $\pt \gtrsim 2.5$ \GeV
with resolution $\sigma(\pt)/\pt$ better than 3\% in the momentum
range of interest ($2.5$--$100$~\GeV).

\section{Event selection}
The data were collected using a single-muon trigger that required transverse momentum $\pt > 18$~\GeV.
Muon candidates are reconstructed either by combining tracks found
separately in the ID and MS (``combined'' muons) or by extrapolating 
ID tracks to include hits in the MS (``segment-tagged'' muons).  Muon tracks are required to satisfy $|\eta| < 2.5$ so as to lie within the angular acceptance of both the ID and MS. 
Muon candidates with $\pt > 3.5$ ($2.5$)~\GeV for $|\eta| < 1.3$
($> 1.3$) are considered.  
The candidate hard scattering $pp$ collision vertex is chosen as the
reconstructed vertex with the highest $\sum \pt^2$ of associated
tracks and the point of closest approach of muon candidate tracks to this
vertex is required to be within 10~mm along the beam axis ($z$).  

Events are required to have at least three identified muons to be
considered for this analysis. One pair of oppositely charged muons is
required to form a \jpsi\ candidate, while an additional muon must
combine with the event's missing transverse momentum \met\ to form a \wpm
candidate. The \wpm\ decay muon is required to match
the muon reconstructed by the trigger algorithm used to collect the
events. The momentum imbalance, \met, is caused by particles, such as
neutrinos escaping detection, detector effects, or unaccounted physics
processes~\cite{Aad:2012re}. The \met\ is calculated as the negative of
the vector sum of the transverse momentum of all reconstructed physics
objects in the event, as well as all calorimeter energy clusters
within $|\eta|<4.9$ not associated with these objects.

At least one of the muons forming the \jpsi\ candidate must have $\pt > 4$~\GeV and at least one must be a
``combined'' inner detector plus muon spectrometer muon.  A vertex fit is performed to constrain the two \jpsi\ muons to originate from a common point. 
The invariant mass of the dimuon system calculated with track
parameters modified by the vertex fit must satisfy $2.5
< m_{\mu^+\mu^-} < 3.5$~\GeV.  The \jpsi\ candidate is required to have transverse momentum \ptjpsi and rapidity $\yjpsi \equiv \tanh^{-1} (p_z^{\jpsi}/E^{\jpsi})$  satisfying \ptjpsi $> 8.5$~\GeV  and 
$|\yjpsi| < 2.1$; this ensures high acceptance and efficiency for the
\jpsi\ candidate.  In addition, as the signal is expected to decrease
faster than background with increasing \ptjpsi, $\ptjpsi < 30$~\GeV is
required to improve the signal-to-background ratio.

The remaining ``combined'' muons are considered as \wpm\ boson decay candidates.  These are required to have $\pt > 25$~\GeV and $|\eta| < 2.4$ in order to be 
within the acceptance of the trigger.  The closest distance of the
\wpm\ decay muon track to the primary vertex must be within 1~mm in $z$.  In order to reduce contamination from 
non-prompt muons produced in heavy flavour decays, the
\wpm\ decay muon is required to have a transverse impact parameter significance
$d_0/\sigma(d_0) < 3$, and it must be isolated. The impact
parameter $d_0$ is defined as the distance of closest
approach of the muon helix to the primary vertex in the $xy$-plane,
and $\sigma(d_0)$ is the expected resolution on the measured
value. The calorimetric and track 
isolation variables are defined as the sums of calorimeter cell \ET\ and track \pt, respectively, within a cone of size $\sqrt{(\Delta \eta) ^2 + (\Delta \phi)^2} = 0.3$ around the
muon direction.  The energy deposit of the muon is not included in the
calorimetric isolation and only tracks (excluding the muon itself) compatible with originating at
the primary vertex and with $\pt > 1$~\GeV
are considered for the track isolation.  
A correction depending on the number of reconstructed vertices is made to the calorimetric isolation to account 
for energy deposits originating from additional proton--proton
collisions that occur in the same bunch crossing.  For the muon
to be considered isolated, the two isolation variables must both be less than
5\% of the muon \pt.  This tight selection discriminates against
the multi-jet background.

In order to select events with a \wpm\ boson, \met\ must exceed
20~\GeV, and the \wpm\ boson transverse mass \WTM\ must exceed
40~\GeV. The \wpm\ boson transverse mass is defined as
\begin{equation*}
\WTM\equiv\sqrt{2\pt(\mu)\met \left(1-\cos(\phi^{\mu}-\phi^{\nu_{\mu}})\right)}
\end{equation*}
where $\phi^{\mu}$ and $\phi^{\nu_{\mu}}$ represent the azimuthal angles
of the muon from the \wpm\ boson decay and the missing transverse
momentum vector, respectively.  If the invariant mass of the \wpm\ muon
 candidate and the \jpsi\ candidate's oppositely charged muon is
 within 10~\GeV of the \Zzero\ boson mass, the event is vetoed.  
After the full selection, 149 events remain, 78 with $|y_{\jpsi}|\le 1$
and 71 with $1<|y_{\jpsi}|<2.1$.  

Prompt \jpsi\ candidates are distinguished from those originating from
$b$-hadron decays through the separation of the primary vertex and the
\jpsi\ decay vertex.  The pseudo-proper time 

\begin{equation*}
\tau \equiv \frac{\vec{L} \cdot \vec{p}_{\mathrm{T}}^{\jpsi} } {\ptjpsi} \cdot \frac{m_{\mumu}}{\ptjpsi}, 
\end{equation*}
is used, where $\vec{L}$ is the separation vector from the primary
vertex to the \jpsi\ decay vertex and $m_{\mumu}$ is the invariant mass of the \jpsi\ candidate.  Prompt \jpsi\ decays have a
pseudo-proper time consistent with zero within resolution.  Non-prompt \jpsi\ decay vertices are displaced from the primary vertex and have positive pseudo-proper
time on average apart from resolution effects.

\section{Signal extraction}~\label{sec:fit}
The two-dimensional \jpsi\ candidate mass and pseudo-proper time scatter distribution is shown
in figure~\ref{fig:masstau}. These data are fit as follows: first, the
two-dimensional fit described below is applied to separate the prompt \jpsi\ component from the
non-prompt \jpsi\ component and combinatorial background.  This is followed by a
fit to the $W$ boson transverse mass \WTM 
to determine the contributions of \wpm\ and multi-jet
background produced in association with a prompt \jpsi.

\begin{figure}[tbp]
  \begin{center}
    \includegraphics[width=0.64\columnwidth]{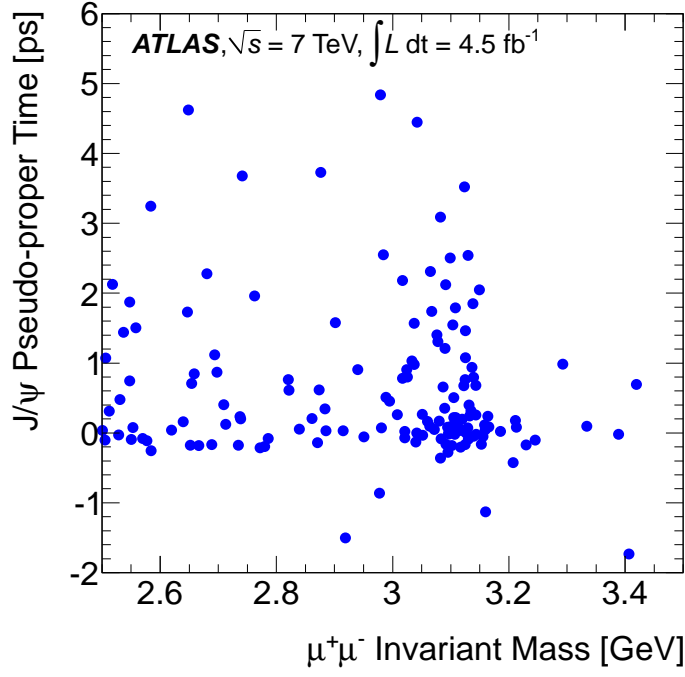}
    \caption{Two-dimensional plot of \wjpsi candidates in
      pseudo-proper time versus $\mu^+\mu^-$ invariant mass in the
      considered region of \jpsi\ rapidity ($|y_{\jpsi}| < 2.1$) and
      transverse momentum ($8.5 < \ptjpsi < 30$~\GeV). Many candidates fall near the \jpsi\ mass of 3.097~\GeV\,\cite{Beringer:1900zz} and pseudo-proper time near $0$~ps, as expected from prompt \jpsi\ production.
      \label{fig:masstau}
    }
  \end{center}
\end{figure}

An unbinned maximum likelihood fit in \jpsi\ candidate invariant mass
and pseudo-proper time is used to obtain yields for prompt
\jpsi, non-prompt \jpsi, and prompt/non-prompt combinatoric
backgrounds. In the dimuon mass variable, the probability density
functions are a Gaussian distribution for the \jpsi\ signal and exponential functions
for the combinatorial backgrounds. For the pseudo-proper time
distribution, the prompt \jpsi\ and prompt combinatoric background components are modelled by the sum of a delta-function distribution and a double-sided
exponential function convolved with a Gaussian resolution function, while the non-prompt \jpsi\ and non-prompt combinatorial background components are modelled by an
exponential function (truncated to zero for $\tau < 0$) convolved with a Gaussian resolution function. The parameters that set the shapes of the fit functions, such as the mass and pseudo-proper time resolution, are considered nuisance parameters and are constrained as described below. The functional forms of the probability density functions are:
\begin{eqnarray*}
M_{\Jpsi}(m_{\mumu}) &=& G(m_{\mumu}; m_{\jpsi}^\textrm{PDG}, \sigma_m) \\
T_{\textrm{prompt \Jpsi}}(\tau) &=& G(\tau; 0, \sigma_\tau) \otimes \left( (1-a) \delta(\tau) + a C_0 \textrm{e}^{-|\tau|/\tau_{\textrm{0}}}\right)\\
T_{\textrm{non-prompt \Jpsi}}(\tau) &=& G(\tau; 0, \sigma_\tau) \otimes \left(C_1 \theta(\tau) \textrm{e}^{-\tau/\tau_{\textrm{1}}}\right)\\
M_{\textrm{prompt bkg}}(m_{\mumu}) &=& C_2 \textrm{e}^{-m_{\mumu}/k_{\textrm{0}}}
\\
M_{\textrm{non-prompt bkg}}(m_{\mumu}) &=& C_3 \textrm{e}^{-m_{\mumu}/k_{\textrm{1}}}
\\
T_{\textrm{prompt bkg}}(\tau) &=&  G(\tau; 0, \sigma_\tau)  \otimes \left( (1-b) \delta(\tau) + b C_4 \textrm{e}^{-|\tau|/\tau_0} \right)
\\
T_{\textrm{non-prompt bkg}}(\tau) &=&  G(\tau; 0, \sigma_\tau) \otimes \left(C_5 \theta(\tau) \textrm{e}^{-\tau/\tau_2}\right).
\end{eqnarray*}
In the above, $G(x; \mu, \sigma)$ is a Gaussian function of $x$ with
mean $\mu$ and width $\sigma$; $\delta$ is the Dirac delta function;
$\theta$ is the step function; $a$, $b$, and the $k_i$ and $\tau_i$
are shape parameters, while the $C_i$ are appropriate normalization constants. The combined
probability density function used for the fit is:
\begin{eqnarray*}
p & \propto & N_{\textrm{prompt \jpsi}} \times M_{\Jpsi}(m_{\mumu}) \times
T_{\textrm{prompt \Jpsi}}(\tau) \\ 
&+& N_{\textrm{non-prompt \jpsi}} \times
M_{\Jpsi}(m_{\mumu}) \times T_{\textrm{non-prompt \Jpsi}}(\tau) \\ 
&+& N_{\textrm{prompt bkg}} \times M_{\textrm{prompt bkg}}(m_{\mumu})  \times T_{\textrm{prompt bkg}}(\tau) \\ 
&+& N_{\textrm{non-prompt bkg}} \times M_{\textrm{non-prompt bkg}}(m_{\mumu}) \times T_{\textrm{non-prompt bkg}}(\tau).
\end{eqnarray*}

To account for the differences in resolution between the central and
endcap detector regions, the fit is performed separately in two regions in
\jpsi\ rapidity, $|y_{\jpsi}| \le 1$ and $1 < |y_{\jpsi}| <
2.1$. The fit is initially made on a large inclusive sample of
\jpsi\ events, selected with the same procedure as the \jpsi\ candidates in this
analysis, and the results are used to constrain the nuisance parameters $\mu$, $\sigma$, $a$, $b$, $k_i$ and $\tau_i$. The
central values and uncertainties on the nuisance parameters from the
inclusive fits are translated into Gaussian constraints on the
parameters during the fit to the \wjpsi candidates.  The values of the
nuisance parameters estimated from the \wpjpsi fits agree well with those determined in the inclusive \jpsi\ sample,
and the resulting uncertainties are similar to the uncertainty
represented by the constraints.  Figure~\ref{subfig:a} shows the mass fit in the full
rapidity region  ($|y_{\jpsi}|<2.1$), whereas in figure~\ref{subfig:b}
the pseudo-proper time fit in the \jpsi\ mass peak region ($3.0<
m(\mu^+\mu^-) <3.2$~\GeV)  is shown. Each projection shows the sum of
two fits, performed separately in two regions in \jpsi\ rapidity: $|y_{\jpsi}| \le 1$ and $1 < |y_{\jpsi}| <2.1$.
\begin{figure*}[tbp]
  \begin{center}
\subfigure[]{\label{subfig:a}    \includegraphics[width=0.49\textwidth]{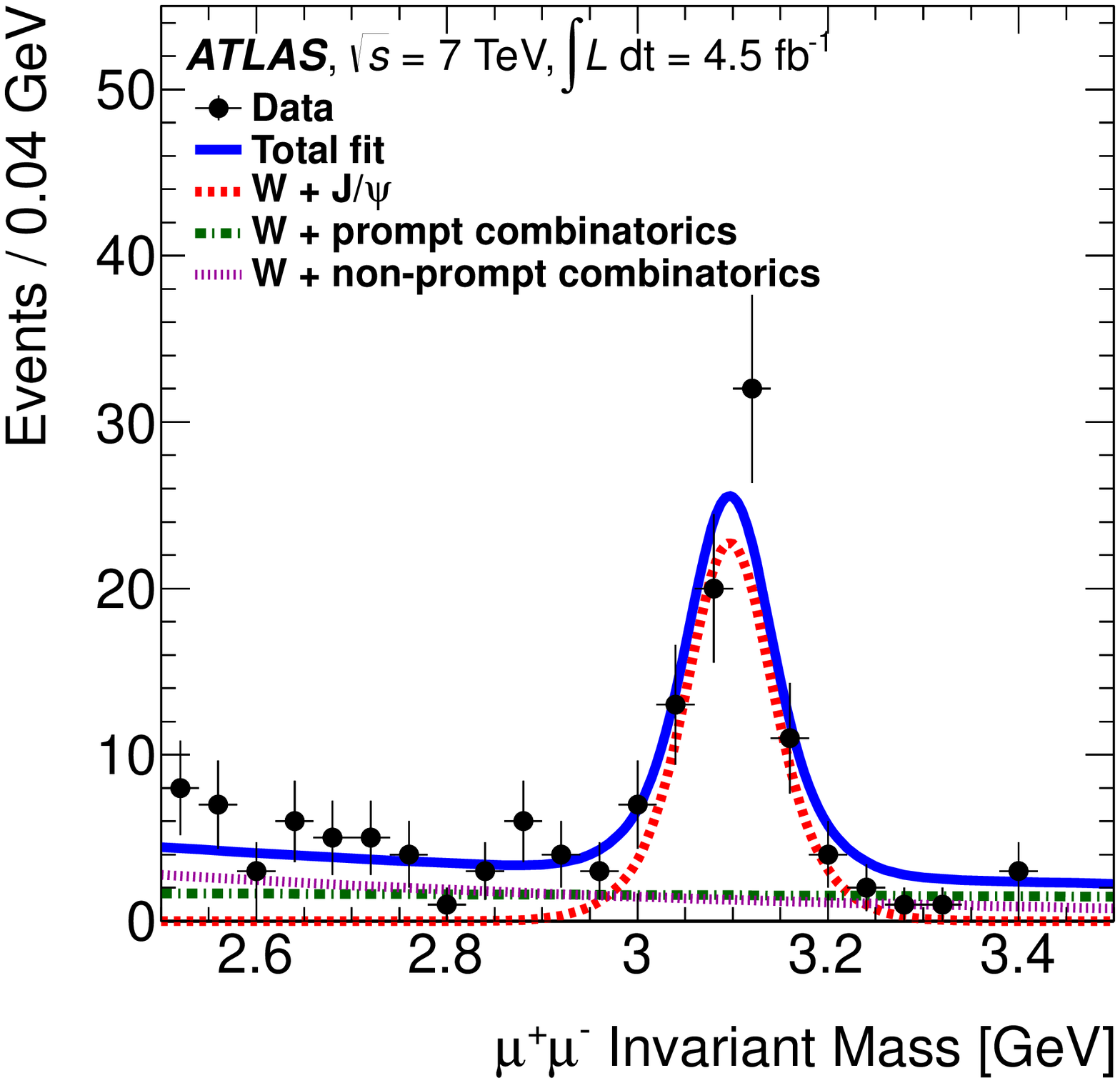}}
\subfigure[]{\label{subfig:b}    \includegraphics[width=0.49\textwidth]{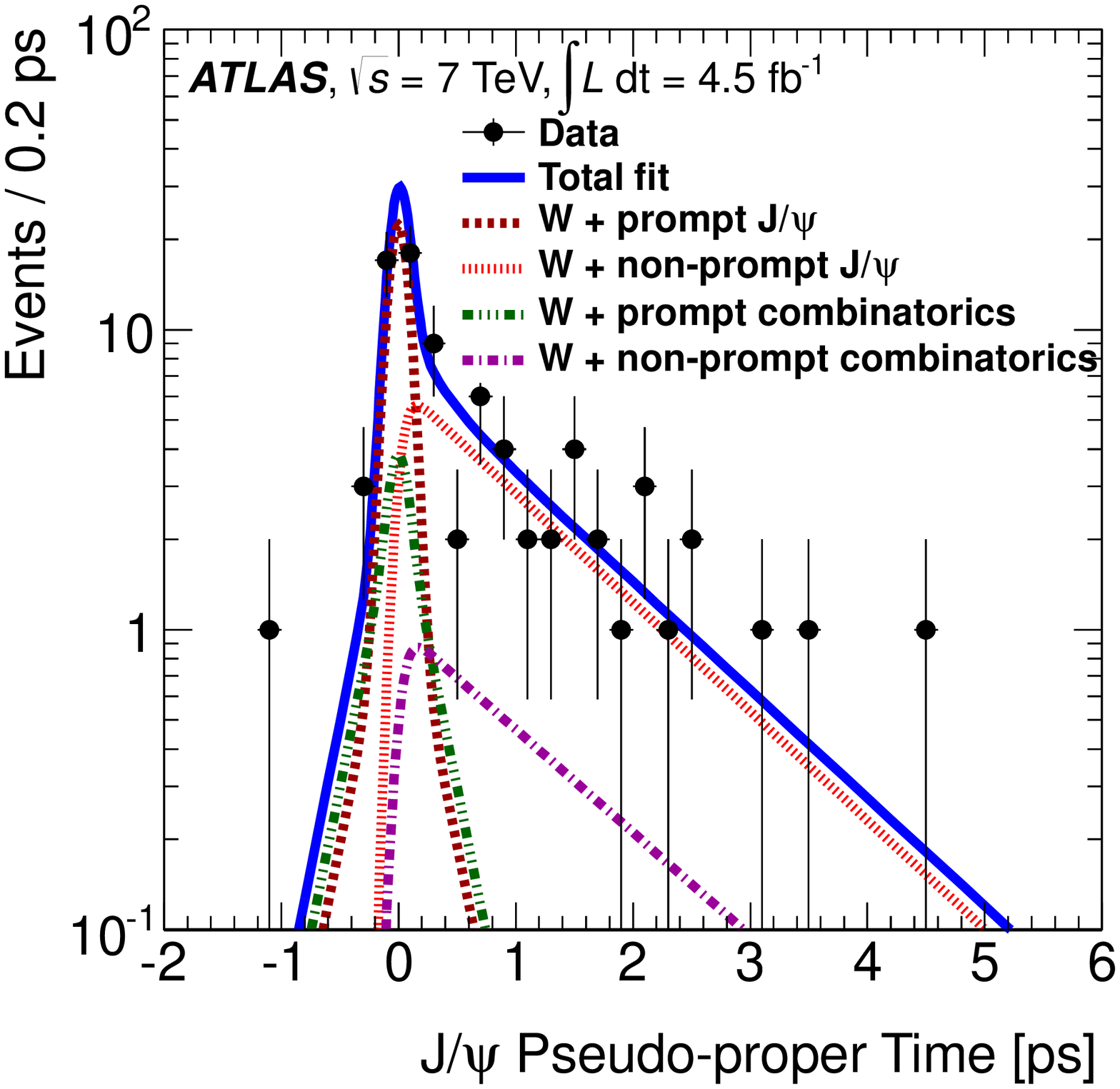}}
\caption{Projections in (a) invariant mass and (b) pseudo-proper time of the 
      two-dimensional mass--pseudo-proper time fit used to extract the prompt $J/\psi$
      candidates in the full rapidity region ($|y_{\jpsi}|<2.1$). The pseudo-proper time distribution is shown for the \jpsi\ mass peak
      region ($3.0< m(\mu^+\mu^-) <3.2$~\GeV). Each projection shows
      the sum of two fits, performed separately in two regions in
      \jpsi\ rapidity $|y_{\jpsi}| \le 1$ and $1 < |y_{\jpsi}| <2.1$. \label{fig:fitmmmb}
    }
  \end{center}
\end{figure*}
The maximum likelihood fit permits the use of the \texttt{sPlot}
procedure\,\cite{Pivk} to assign weights to each event for each component of
the total probability density function (prompt \jpsi, non-prompt \jpsi, prompt
combinatoric background, non-prompt combinatoric background).  These weights
can be used to determine the spectra of variables in the prompt \jpsi\ signal,
removing the contribution from the other sources.
The procedure functions as a background subtraction technique that takes
advantage of the full information available from the fit, and is used to
obtain various kinematic distributions.

The robustness of the fit was tested by varying the signal and background
parametrizations; changes of at most a few percent in the prompt \jpsi\ yield
were observed and are taken as systematic uncertainties (see
Section~\ref{sec:results}).  Repeated fits to an ensemble of
pseudo-experiments generated with Poisson-distributed yields and
Gaussian-distributed nuisance parameters showed that the fit has no
significant statistical biases, and that the uncertainties reported by the fit
are accurate.

\section{Backgrounds}
A number of possible background contributions to \wpjpsi production are considered. 
Production of $\wpm$ bosons in association with $b$ quarks, with
subsequent $b$-hadron decay to \jpsi\ and other particles, produces the
desired signature, except that the \jpsi\ mesons will not be prompt.
The fit will categorize these events into the non-prompt \jpsi\ background category and the effect on the prompt signal yield is estimated to be negligible.
From simulated \ttbar\ events the prompt yield in our dataset is predicted to be less than 0.28 events at 95\% credibility level. The ratio of $\wpm + b$ to \wpm\ production
from ATLAS measurements\,\cite{PhysRevD.85.072004,Aad2012418} is found
to be consistent (albeit with large uncertainties) with the ratio of \wpm + non-prompt \jpsi\ to $\wpm$ production in
this analysis, when the differing phase spaces of the analyses are considered. 

Decays of $B_c\to\jpsi \mu^{\pm} \nu_{\mu} X$ produce a \jpsi and an
additional muon. As the $B_c$ lifetime is shorter than that of other
weakly-decaying $B$ hadrons, these \jpsi may be mistakenly taken as
prompt candidates. No candidate events are found to have a three-muon
invariant mass below 12~\GeV, which is far above the $B_c$ mass of
6.277~\GeV~\cite{Beringer:1900zz}, hence this background is
negligible.

The production of $Z$ bosons, followed by the decay \zmumu, can
produce the signal signature if an additional muon candidate and \met\
due to jet mismeasurements or neutrinos are found.  This background is eliminated by vetoing
events where a pairing of oppositely charged muons has an invariant mass within  10~\GeV of the $Z$ boson mass.

Multi-jet production, especially of heavy-quark jets, can produce
candidates with multiple reconstructed muons and \met\, due to either
neutrinos or mismeasurement of the jet energy.  
This contribution is separated from the real \wjpsi component of the observed events using the \wpm\ boson transverse mass \WTM\ as a discriminating variable.
The \WTM\ distribution of events weighted by
signal (prompt \jpsi) \texttt{sPlot} weight is fit
to a sum of a multi-jet template and a \wpm\ boson signal template
(the templates are shown in figure~\ref{subfig:WmT_template}). 
\begin{figure*}[tbp]
  \begin{center}
\subfigure[]{\label{subfig:WmT_template}     \includegraphics[width=0.49\columnwidth]{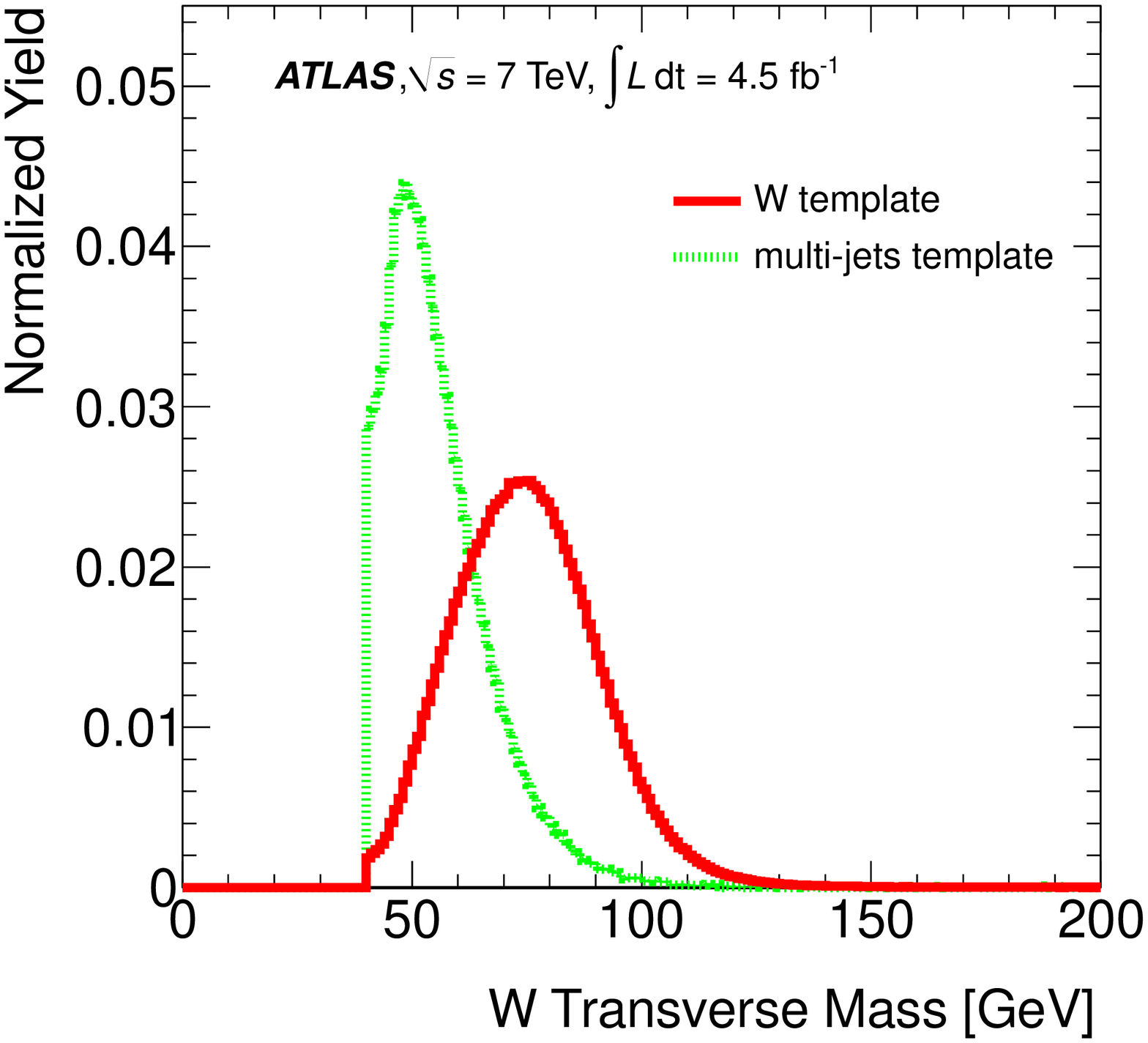}}
\subfigure[]{\label{subfig:WmT_fit}     \includegraphics[width=0.49\columnwidth]{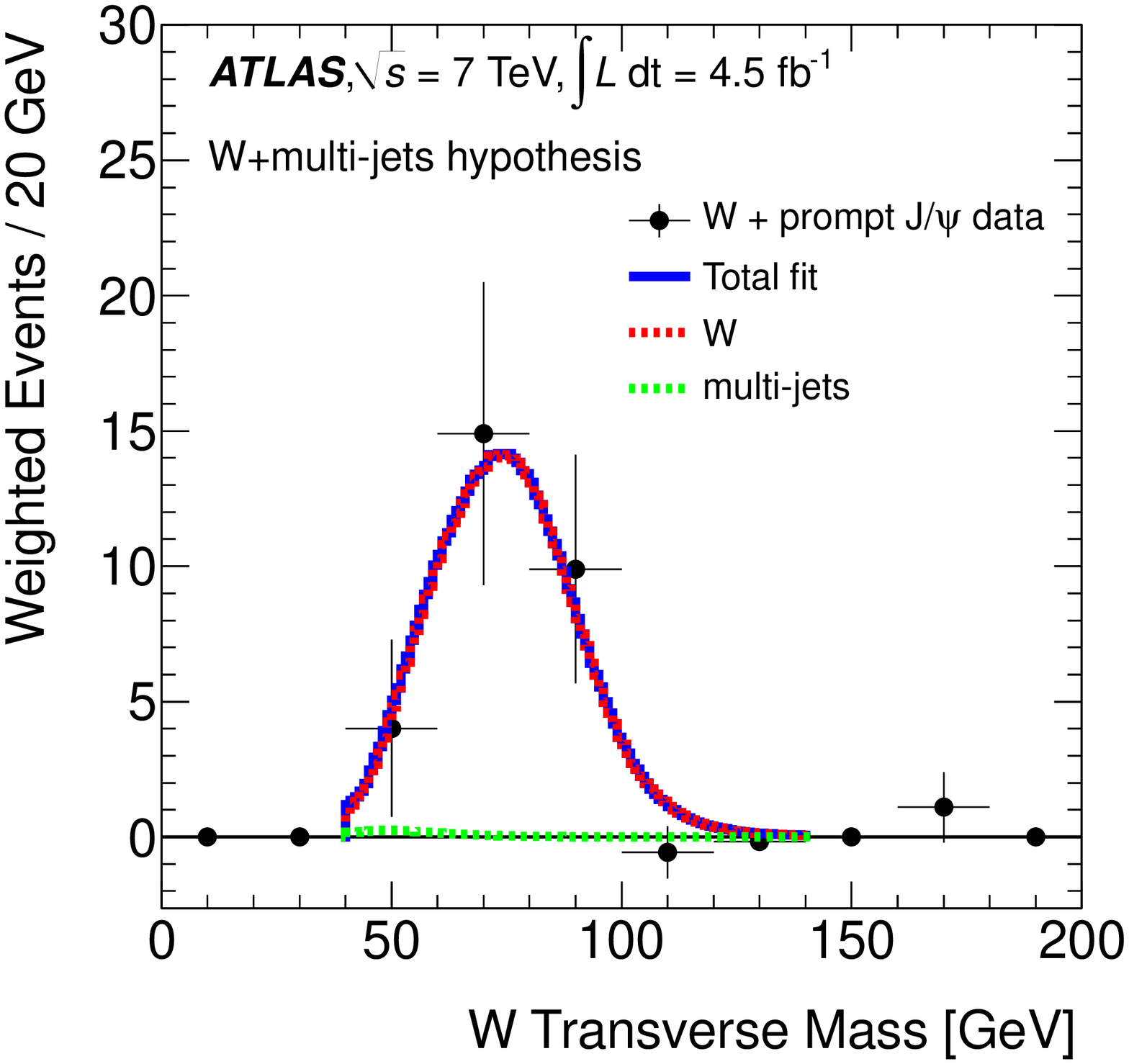}}
\caption{(a) Unit-normalized templates for $W$ boson transverse mass \WTM\ for multi-jet background and \wpm\ boson signal.
      (b) \texttt{sPlot}-weighted $W$ boson transverse mass distribution for \wpjpsi candidate events
      with a fit to the \wpm\ boson and multi-jet components. The fit is performed in the region 40--140~\GeV in \WTM.
      \label{fig:WmT}
    }
  \end{center}
\end{figure*}
The multi-jet background shape in \WTM\ is estimated using the
distribution in events with non-isolated muons, which are dominated by
multi-jet production. The \wpm\ template is obtained
from Monte Carlo simulation.  Events were produced by the \textsc{Alpgen} event generator
\cite{Mangano:2002ea}, interfaced to \textsc{Herwig}
\cite{Corcella:2000bw} for parton showers and hadronization, and
\textsc{Jimmy}\,\cite{Butterworth:1996zw} for simulation of the
underlying event. The detector response is modelled using the \textsc{Geant}4-based ATLAS full simulation framework\,\cite{Agostinelli:2002hh,Aad:2010ah}.

The total yield for prompt \jpsi\ production,
shown in Table~\ref{table:yields}, is \origsignalyield\ events. 
The result of the $\chi^2$ fit is shown in figure~\ref{subfig:WmT_fit}.  
The estimate from the \WTM\ fit is that there are $0.1 \pm 4.6$ multi-jet
events in the sample, providing strong support for the hypothesis
that the sample is dominated by \wpjpsi events.
The fraction of multi-jet events is smaller than 0.31 at $95\%$ credibility level.

The probability that a \wpm\ candidate and a \jpsi\ candidate are
produced in different proton--proton collisions that occur in the same
bunch crossing (``pileup'') is
estimated as follows. Given the beam conditions of the dataset, the
mean number of extra collisions within 10~mm of the primary vertex is
calculated to be $N_\mathrm{extra} = 0.81 \pm 0.08$; this value is
computed from the mean number of collisions per proton--proton bunch
crossing $\mu$ and the geometric parameters of the interaction
region.  
Here $\mu$ is defined as $\mu = \mathcal{L} 
\sigma_{\textrm{inel}}/n_b f_r$,
with $\mathcal{L}$ being the luminosity, $n_b$ the number of colliding
bunch pairs, $f_r$ the accelerator revolution frequency, and
$\sigma_{\textrm{inel}}$ the $pp$ inelastic cross section,
assumed to be equal to  71.5 mb\,\cite{Aad:2011dr}.  
For a $(|y_{\Jpsi}|, \pt^{\Jpsi})$-bin, the probability for a \jpsi\ to be produced in a $pp$ collision in that kinematic bin is determined as

\begin{equation*}
 P_{\Jpsi} = \frac{\sigma_{\jpsi}^\mathrm{bin}}{\sigma_{\textrm{inel}}} = \frac{1}{\sigma_{\textrm{inel}}} \int_{\textrm{bin }} \frac{\mathrm{d}^{2}\sigma(pp \to \jpsi\ X)}{\mathrm{d}y\ \mathrm{d}\pt} \mathrm{d}y\ \mathrm{d}\pt
\end{equation*}
using the double-differential \jpsi\ production cross sections as
measured\,\cite{Aad:2011sp} at $\sqrt{s}=7$~TeV.

 Since $\mathcal{L}$ is determined independently from $\sigma_{\textrm{inel}}$ using van der Meer scan calibration\,\cite{Aad:2013ucp}, 
$\sigma_{\textrm{inel}}$ is a proportionality factor between $\mu$ 
and $\mathcal{L}$, and therefore $N_\mathrm{extra} \propto \sigma_{\textrm{inel}}$.
As a result the dependence on $\sigma_{\textrm{inel}}$ cancels in the overlap 
probability 
$N_\mathrm{extra} P_{\Jpsi}$. 
Multiplying the overlap probability by the number of \wpm\
candidates in the fiducial region, $N_\mathrm{pileup}=N_\mathrm{extra} P_{\Jpsi}
\mathcal{L} \sigma_{W^{\pm}}$,
yields an estimated total of $1.8\pm0.2$ for such pileup overlap events in
the sample.  This
background is subtracted when the cross-section ratios are calculated. 
 
\section{Double parton scattering}
It is possible for the \wpm\ and \jpsi\ to originate from two different parton interactions in the same proton--proton
collision, in a double parton scattering process.  The
standard ansatz~\cite{Paver:1982yp} is adopted that, for a collision in which a hard process (here \wpm\ production) occurs, the probability
of an additional (distinguishable) process (here prompt $\jpsi$ production) is parameterized as 

\begin{equation*}
P_{\jpsi\,|\,\Wpm} = \sigma_{\jpsi}/\sigma_{\textrm{eff}}. \tag{1}
\label{eq:DPS}
\end{equation*}
The effective area parameter $\sigma_{\textrm{eff}}$ accounts for the
geometric size of the proton and transverse parton correlations, and is
assumed to be independent of the scattering process.  It is taken
to be $15 \pm 3 \stat ^{+5}_{-3}\syst$~mb as measured using
$\wpm\to \ell \nu_{\ell} + 2-$jet events\,\cite{Aad:2013bjm}.  The two interactions are treated as independent and uncorrelated.
This procedure relies upon an assumption of factorization between the longitudinal and transverse components of the double parton distribution functions\,\cite{Blok:2010ge,Diehl:2011tt,Gaunt:2011xd,Bartalini:2011jp}.
Recent studies\,\cite{Blok:2011bu} highlight that this assumption must fail at some level, but the current data is not sufficiently precise to allow for investigation of such effects.
Studies of process and scale dependence\,\cite{Blok:2013bpa} of $\sigma_{\textrm{eff}}$ indicate that the uncertainty on the value used here covers variation due to these effects.
The prompt \jpsi\ cross section from the ATLAS measurement~\cite{Aad:2011sp} is used, as in the pileup estimation,
corrected to the fiducial phase space of this measurement, and accounting for spin-alignment uncertainties that result from the correction.
The total number of DPS events in the signal yield is estimated to be $10.8\pm4.2$ events. 

A uniform distribution in the azimuthal angle between the \wpm\ and
\jpsi\ momenta is expected from DPS, under the assumption that the two
interactions are independent. The flat DPS template is verified using \textsc{pythia8}\,\cite{Sjostrand:2007gs} Monte Carlo simulation. Simulations in both the colour-singlet and colour-octet models
predict a distribution strongly peaked near $\Delta\phi=\mathrm{\pi}$ for the SPS contribution (with the exact shape dependent on kinematics and the relative size of the underlying contributions).
The \texttt{sPlot}-weighted distribution of this variable for the data  has a peak
near $\mathrm{\pi}$ and a tail extending towards zero, as shown in figure~\ref{fig:deltaphi}.  
This suggests that the observed \wpjpsi\ candidates include both SPS
and DPS events.  However, this distribution is not used to separate SPS and DPS events.
Overlaid on figure~\ref{fig:deltaphi} is an estimate of the DPS rate using the previously described ansatz. The determined rate is compatible with the size of the 
flat component of the observed $\Delta\phi$ distribution.

\begin{figure}[tbp]
  \begin{center}
\includegraphics[width=0.64\columnwidth]{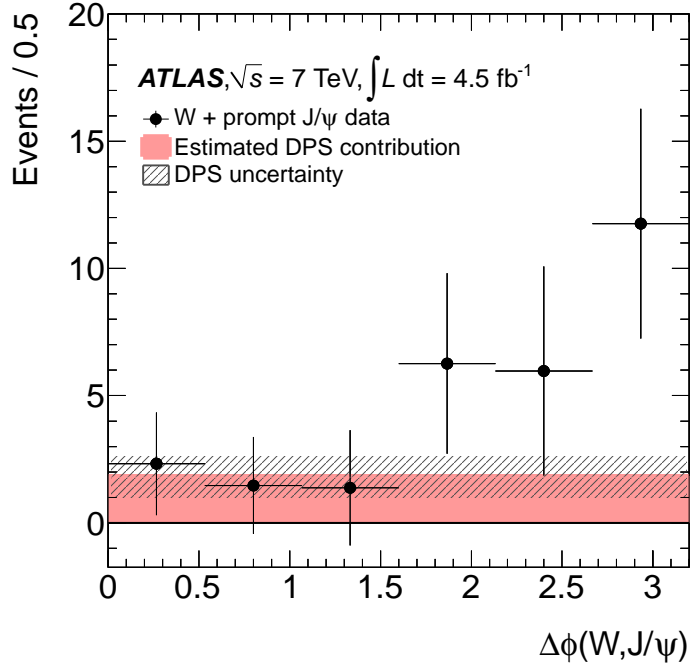}
    \caption{The \texttt{sPlot}-weighted azimuthal angle between the
      \wpm\ and the \jpsi\ is shown for \wpjpsi candidates. No
      efficiency or acceptance corrections are applied. 
      The determined DPS contribution (with systematic uncertainties)
      is overlaid, using a flat DPS template
      validated using \textsc{pythia8}\,\cite{Sjostrand:2007gs} Monte Carlo simulation and normalized to the total rate as estimated using the DPS ansatz in \protect{eq.~\ref{eq:DPS}}.  The hashed region shows the uncertainty on the DPS estimate.
      \label{fig:deltaphi}
    }
  \end{center}
\end{figure}
\section{Results}
\label{sec:results}
Table~\ref{table:yields} shows the yields resulting from the two-dimensional fit in each of the two detector regions, barrel ($|\yjpsi|\le 1.0$) and endcap ($1 <|\yjpsi|<2.1$), 
including the statistical uncertainty from the fit for the prompt \jpsi, non-prompt \jpsi, and combinatoric
background components.  The signal significance is calculated using pseudo-experiments in which events conforming with the
 background-only hypothesis are generated and then fit with the background-only hypothesis and the signal+background hypothesis to
 extract the likelihood ratio of the two hypotheses. The $p$--value
 shown in the Table is the probability for background-only pseudo-experiments to yield a
 likelihood ratio value larger than the value obtained in data. The 
combination of the two \yjpsi regions rejects the background-only hypothesis 
at 
\significance.

\begin{table}[tbp]
\begin{center}
\begin{tabular}{lccc}
\hline \hline
\multicolumn{4}{c}{Yields from two-dimensional fit}\\
\hline
Process & Barrel & Endcap & Total \\
\hline
\hline
Prompt \jpsi\ & $  10.0  {}^{+  4.7  }_{ -4.0  }$ & $  19.2  {}^{+  5.8  }_{ -5.1  } $& $29.2
{}^{+7.5} _{-6.5} (*) $\\
Non-prompt \jpsi\ &  
$  27.9  {}^{+  6.5  }_{ -5.8  }$&$  13.9  {}^{+  5.3  }_{ -4.5  }$& $41.8
{}^{+8.4} _{-7.3} $\\
Prompt background & 
$  20.4  {}^{+  5.9  }_{ -5.1  } $& $  18.8  {}^{+  6.3  }_{ -5.3  } $& $39.2
{}^{+8.6} _{-7.3} $\\
Non-prompt background &
$  19.8  {}^{+  5.8  }_{ -4.9  } $ & $  19.2  {}^{+  6.1  }_{ -5.1  } $& $39.0
{}^{+8.4} _{-7.1} $\\
\hline\hline
$p$-value & $8.0\times 10^{-3}$ & $1.4\times 10^{-6}$ & $2.1\times 10^{-7}$\\
\hline
Significance ($\sigma$) &2.4& 4.7 & 5.1\\
\hline
\hline
\footnotesize{ (*) of which $1.8\pm0.2$ originate from pileup} 
\end{tabular}
\end{center}
\caption{\label{table:yields}The event yields for the prompt \jpsi,
  non-prompt \jpsi, and combinatorial background are shown.
 The errors shown include the statistical uncertainties and the systematic uncertainties from the nuisance parameters of the fit.
The significance, in standard deviations, is
  calculated using the $p$-value from pseudo-experiments.}
\end{table}

To determine the ratio of the \wpjpsi cross section to the \wpm cross section,
the number of inclusive \wpm\ events is determined from data.  A
sample of \wpm\ candidates is formed by selecting all events that satisfy the \wpm\ part of the \wpjpsi\ requirements. The
\Zzero$+\textrm{jets}$, $t\bar t$, and diboson backgrounds to
inclusive \wpm\ production are estimated from Monte Carlo simulation
(\textsc{Alpgen}~\cite{Mangano:2002ea}, \textsc{MC@NLO}~\cite{Frixione:2002} and \textsc{Herwig}~\cite{Corcella:2000bw}, respectively).
The multi-jet contribution is estimated using the same technique as
used for deriving the multi-jet template for the \wpjpsi signal, except
that the normalization is also determined by that method instead of being fit to data.  
The number of \wpm\ boson candidates is found to be $1.48\times
10{}^7$, consistent with next-to-next-to-leading-order (NNLO) pQCD
predictions\,\cite{Gavin:2010az,Gavin:2012sy} taking into account the ATLAS detector performance.

First the fiducial cross-section ratio
\begin{eqnarray*}
\rfid &=& \frac{\mathrm{BR}(\jpsi\rightarrow\mu^+ \mu^-)}{\sigma_\mathrm{fid}(pp \to \Wpm)}\cdot \frac{\mathrm{d}\sigma_\mathrm{fid}(pp \to \Wpm + \Jpsi)}{\mathrm{d}y} \\&=& \frac{N^{\mathrm{ec}}(\Wpm+\Jpsi)}{N(\Wpm)}\frac{1}{\Delta y} - R^\mathrm{fid}_\textrm{pileup},
\end{eqnarray*}
 is defined, where $N^{\mathrm{ec}}(\Wpm+\Jpsi)$ is the yield of
 \wpjpsi events after correction for the \jpsi\ muon reconstruction
 efficiencies, $N(\Wpm)$ is the background-subtracted yield of
 inclusive \wpm\ events, $\Delta y = 4.2$ is the size of the fiducial
 region in \yjpsi, and $R^\mathrm{fid}_\textrm{pileup}$ is the
 expected pileup background contribution in the fiducial \jpsi
 acceptance.  For \rfid, corrections are not applied for the
 incomplete acceptance for \jpsi\ decay muons, nor for the \wpm acceptance.
 Also, it is noted that only the cross section for $8.5 < \ptjpsi < 30$~\GeV is considered.

The statistical uncertainties associated with the fit are calculated by
fixing the nuisance parameters and performing the fit again. When the
nuisance parameters are allowed to float within the Gaussian
constraint, the total uncertainty on each yield is the quadratic sum of the
statistical and systematic uncertainties. 

The \jpsi\ transverse momentum distribution may be different in
inclusive \jpsi\ events and \wpjpsi events. 
Since the fit nuisance parameters from the inclusive fit are used
during the \wpjpsi fit, as described in section~\ref{sec:fit}, this
can affect the extracted yields, due to the different \jpsi\ \pt\
spectrum of the inclusive \jpsi\ and \wjpsi processes. This is
estimated to have an effect of $<1\%$ on the prompt \jpsi\ yields
extracted by the fit, by performing the fits with unconstrained
nuisance parameters in different \jpsi\ \pt\ ranges and comparing the
yields with those from the constrained fits in the nominal \jpsi\ \pt\
range. Changing the functional forms of the fit, for
example by changing the single Gaussian to a double Gaussian for the
signal, or the exponential function to a polynomial function for the
background, results in changes in the yields of up to $4\%$, which is taken as a systematic
uncertainty.

The efficiency and acceptance of the \wpm\ boson are assumed to be the
same for inclusive \wpm\ events and \wjpsi events when calculating the ratio. The
uncertainty due to this assumption is estimated by reweighting the \wpm\ transverse
momentum spectrum in the simulated inclusive sample to match the
observed spectrum from \wjpsi\ events and noting the change in
efficiency, which is $(2$--$5)\%$. The low--\pt\ muon efficiencies are determined from
data and cross-checked with efficiencies measured from \jpsi\
simulation using \textsc{pythia8}\,\cite{Sjostrand:2007gs}, with the
difference interpreted as a systematic uncertainty of $(3$--$5)\%$, while the uncertainty due to
the muon momentum scale is found to be negligible. 

In addition to reporting \rfid, results are presented after being corrected for the
fiducial acceptance of the muons from the \jpsi\ decay, but maintaining
the \jpsi\ \pt\ (8.5 -- 30 ~\GeV) and rapidity (-2.1 -- 2.1) range:  \begin{eqnarray*}
\rincl &=& \frac{\mathrm{BR}(\jpsi\rightarrow\mu^+ \mu^-)}{\sigma_\mathrm{fid}(pp \to \Wpm)}\cdot \frac{\mathrm{d}\sigma(pp \to \Wpm + \Jpsi)}{\mathrm{d}y} \\
&=& \frac{N^{\mathrm{ec+ac}}(\Wpm+\Jpsi)}{N(\Wpm)}\frac{1}{\Delta y} - R_\textrm{pileup},
\end{eqnarray*}
where $N^{\mathrm{ec+ac}}(\Wpm+\Jpsi)$ is the yield of \wpjpsi\ events
after \jpsi\ acceptance corrections and efficiency corrections for both
\jpsi\ decay muons, $R_\textrm{pileup}$ is the expected pileup contribution in the full \jpsi\ decay phase space, and other variables are as for \rfid.
The \jpsi\ spin-alignment, which determines the angular distribution of the muons from the \jpsi\ decay and thus 
modifies the acceptance for the \jpsi\ to be detected within the fiducial volume, is not known and is dependent on the underlying production mechanism.
Five extreme scenarios that bound the possible variation are
considered for the acceptance and the difference is assigned as a systematic
uncertainty. These scenarios (isotropic, longitudinal,
transverse$+$, transverse$-$ and  transverse$0$) depend on the \jpsi\ spin-alignment
angles, the angle between the direction of the positive muon momentum
in the \jpsi\ decay frame and the \jpsi\ line of flight, and the angle
between the \jpsi\ production plane and the decay plane formed by the
direction of the \jpsi\ and the positive muon~\cite{Aad:2011sp}. 

Table~\ref{tab:syst} summarizes the main sources of uncertainties for
this analysis. Other uncertainties related to the \met\ scale,
luminosity, and the \wpm\ decay muon cancel in the ratio. 

\begin{table}[hbtp]
\begin{center}
\begin{tabular}{lccc}
\hline\hline

Source & Barrel & Endcap\\
\hline
\jpsi\ muon efficiency & $(3$--$5)\%$ & $(3$--$5)\%$\\
\wpm\ boson kinematics & $2 \%$ & $5 \%$\\
Fit procedure & ${}^{+3}_{-2}\%$ & ${}^{+2}_{-1}\%$\\
Choice of fit nuisance parameters & $1\%$ & $1\%$\\
Choice of fit functional forms & $4\%$ & $4\%$\\
Muon momentum scale & \multicolumn{2}{c}{negligible}\\
\hline
\jpsi\ spin-alignment & ${}^{+36} _{-25} \%$ & ${}^{+27}_{-13} \%$\\
\hline
Statistical &${}^{+47}_{-40}\%$ & ${}^{+30}_{-27}\%$\\
\hline
\hline
\end{tabular}
\end{center}
\caption{\label{tab:syst}Summary of the main sources of
  uncertainty for the measurements of \rfid and \rincl; systematic, spin-alignment and
  statistical uncertainties are shown. Only uncertainties that do not
  cancel in the ratio of \wpjpsi to \wpm\ rates are included. The
  spin-alignment uncertainty is not present for the \rfid measurement.}
\end{table}

The isotropic spin-alignment scenario is assumed for the central
result, and the variations of the result with the different spin-alignment scenarios are also reported in \textsc{hepdata}\,\cite{hepdata}.
The DPS contribution to \rincl is estimated to be $(48 \pm 19) \times 10^{-8}$. 
Subtracting the DPS contribution from \rincl gives an estimate \rdpssub of the single parton scattering rate, which can be directly 
compared with leading-order (LO) colour-singlet pQCD predictions\,\cite{Lansberg:2013wva} and next-to-leading-order (NLO) colour-octet predictions\,\cite{Gang2011}.

The values of the three measured ratios are shown in figure~\ref{fig:ratios} and are:
\begin{eqnarray*}
\rfid &=& (51 \pm 13 \pm 4) \times 10^{-8} \\  
\rincl &=& (126 \pm 32 \pm 9  {}^{+41}_{-25}) \times
10^{-8} \\  
\rdpssub &=& (78 \pm 32 \pm 22 {}^{+41}_{-25})  \times 10^{-8}, 
\end{eqnarray*}
where the first uncertainty is statistical, the second is systematic
and the third (where applicable) is the uncertainty due to
spin-alignment. The systematic uncertainty on the DPS-subtracted ratio includes
the uncertainty on the estimated DPS contribution, which itself
includes a separate spin-alignment uncertainty.
\begin{figure}[tb]
  \begin{center}
    \includegraphics[width=0.64\columnwidth]{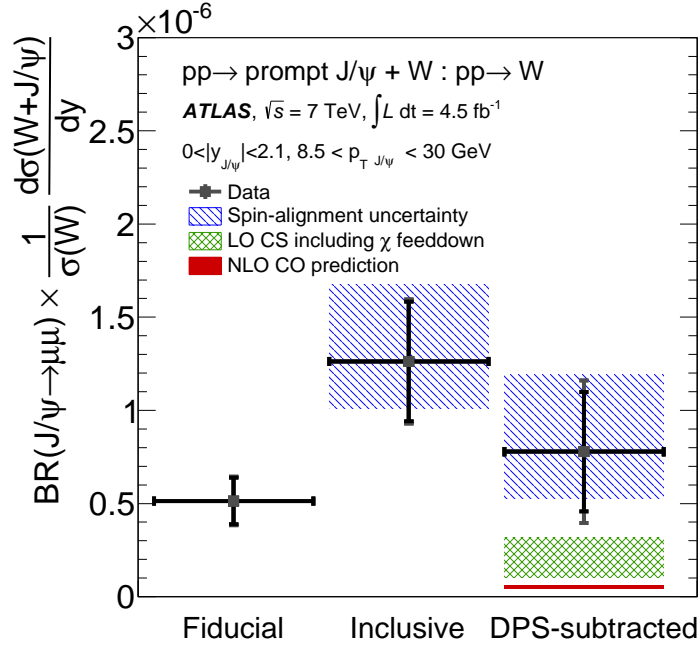}
\caption{The \wpjpsi\ : \w production cross-section ratio
  in the \jpsi\ fiducial region (Fiducial), after correction for \jpsi
  acceptance (Inclusive), 
  and after subtraction of the double parton scattering component (DPS-subtracted). 
  The shaded band represents the envelope of variation due to different possible spin-alignment configurations.
  Inner error bars represent statistical uncertainties, outer error
  bars represent statistical and systematic uncertainties added in
  quadrature.  The LO colour-singlet (CS) and NLO colour-octet (CO)
  predictions for SPS production are shown in comparison.
      \label{fig:ratios}
    }
  \end{center}
\end{figure}

\subsection*{Comparisons with theoretical expectations}
For comparison of the DPS-subtracted ratio to theory, the LO colour-singlet and NLO colour-octet predictions for \wpjpsi\,\cite{Gang2011}
are normalized to NNLO calculations of the \wpm\ production cross
section (5.08 nb), derived from \textsc{fewz 3.1.b2}\,\cite{Gavin:2010az,Gavin:2012sy}.
The expected SPS cross-section ratio \rdpssub from normalized next-to-leading-order colour-octet calculations is $(4.6 $--$ 6.2) \times  10^{-8}$, with the range corresponding to different scales as explained below.
These predictions assume that pure CO contributions dominate this
process, and hence do not include any colour-singlet diagrams.
The renormalization ($\mu_R$) and factorization ($\mu_F$) scales are
set to the \wpm\ boson mass, $\mu_R=\mu_F=m_W$,  
and the NRQCD scale ($\mu_\Lambda$) is taken to be $\mu_\Lambda=m_\textrm{charm}=1.5$~GeV.
Two sets of colour-octet long-distance matrix elements are used as input parameters to the theory, obtained from two different parameter fits to experimental data\,\cite{CO_LDME1, CO_LDME2},
and the variation between the results obtained is assigned as a systematic uncertainty.

The CS model described in\,\cite{Lansberg:2013wva}  includes contributions from $sg\to J/\psi+c+W$ and
$q\bar{q}^\prime\to\gamma^*W\to J/\psi W$, and accounts 
for indirect production in the \wpjpsi\ rate through decays of excited charmonium states of both
$\Wpm + \textrm{direct}~\psi(2S)$ and $\Wpm +
\textrm{direct}~\chi_{cJ}$ production. 
The normalized leading-order colour-singlet model predicts the SPS ratio to be $(10 $--$ 32) \times  10^{-8}$ 
for the phase space region considered in this paper, with the range
corresponding to different scales as explained below. The uncertainties on the colour-singlet prediction arise from  varying
the charm quark mass in the range $m_\textrm{charm}=(1.5\pm 0.1)$~\GeV\
and independently varying the factorization and renormalization scales between $0.75\times\mu_{R,F}$ and $2\times\mu_{R,F}$ of their central value $\mu_R=\mu_F=m_W$.

The leading-order CS contributions are nearly an order of magnitude larger than the next-to-leading-order CO contributions, with the CS prediction
being consistent with the measured DPS-subtracted rate within the
current experimental and theoretical uncertainties. This emphasizes
that far from being a distinctive signature of CO production, this
process appears to be dominated by CS production. This shortfall in the CO production prediction might be explained by the presence of large higher-order contributions\,\cite{CO_LDME1,CO_LDME2} or
a possible breakdown of NRQCD universality\,\cite{Butenschoen:2012qr}.
Figure~\ref{fig:deltaphi} supports the hypothesis of DPS factorization (within current experimental uncertainties), 
but possible modifications to this formalism may also modify the DPS-subtracted rate.
Nonetheless, large uncertainties on the result imply that current
predictions for SPS production are compatible with the measurement at the $2\sigma$ level.

Figure~\ref{fig:total} shows the distribution of the differential
cross-section ratio $\mathrm{d}\rincl/\mathrm{d}\pt$ as a function of \ptjpsi, with an
estimate of the differential DPS contribution. The data suggest
that SPS is the dominant contribution to the total rate at low \jpsi
transverse momenta. The results are also shown in Table~\ref{table:diff}.
\begin{figure}[tb]
  \begin{center}
    \includegraphics[width=0.62\columnwidth]{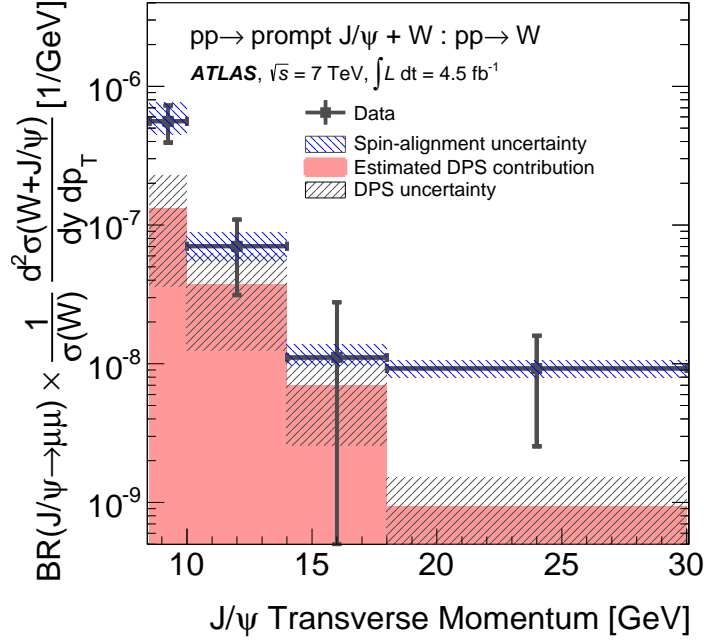}
    \caption{
      The inclusive (SPS+DPS) cross-section ratio $\mathrm{d}\rincl/\mathrm{d}\pt$ is shown as a function of
      \jpsi\ transverse momentum. The shaded uncertainty corresponds to the variations due to the various spin-alignment scenarios. 
      The DPS estimate is overlaid,
with the uncertainty again shown by a shaded region.
      \label{fig:total}
    }
  \end{center}
\end{figure}

\begin{table}[hbtp]
\begin{center}
\begin{tabular}{l l@{$\pm$}l@{$\pm$}l@{${}$}l l@{$\pm$}l}
\hline\hline
$\yjpsi \times \ptjpsi$ Bin & \multicolumn{4}{c}{Inclusive (SPS+DPS) ratio $\mathrm{d}\rincl/\mathrm{d}\pt$ ($\times 10^{-6}$)} & \multicolumn{2}{c}{DPS ($\times 10^{-6}$)} \\ 
\hline
$(0, 2.1) \times (8.5, 10)$ & $ 0.56 $ &   $ 0.16 \textrm{(stat)} $ & $ 0.04 \textrm{(syst)} $ & $^{+ 0.21}_{- 0.11 }\textrm{(spin)} $ &$ 0.13 $ & $ 0.10 $ \\
$(0, 2.1) \times (10, 14)$ &  $ 0.070 $ &  $ 0.039 \textrm{(stat)} $ & $ 0.006 \textrm{(syst)} $ & $ ^{+ 0.019 }_{- 0.016 }\textrm{(spin)} $ &$ 0.04 $ & $ 0.03 $ \\
$(0, 2.1) \times (14, 18)$ &  $ 0.011 $ &  $ 0.017 \textrm{(stat)} $ & $ 0.001 \textrm{(syst)} $ & $^{+ 0.003 }_{- 0.002 }\textrm{(spin)} $ &$ 0.007 $ & $ 0.004 $ \\
$(0, 2.1) \times (18, 30)$ &  $ 0.0092 $ & $ 0.0067 \textrm{(stat)} $ & $ 0.0006 \textrm{(syst)} $ & $^{+  0.0012 }_{- 0.0013 }\textrm{(spin)} $ &$ 0.0009 $ & $ 0.0006 $ \\
\hline
\hline
\end{tabular}
\end{center}
\caption{\label{table:diff}The inclusive (SPS+DPS) cross-section ratio
  $\mathrm{d}\rincl/\mathrm{d}\pt$ as a function of \jpsi\ transverse
  momentum, along with the estimate of the DPS contribution.}
\end{table}

\FloatBarrier
\section{Conclusions}
In summary, the ATLAS Collaboration has observed \wpjpsi production at
\significance\ significance in 4.5~\ifb\ of $\sqrt{s} = 7$ TeV $pp$ collisions at the LHC.
Additionally, the fiducial cross-section ratio of \wpjpsi production
relative to inclusive \wpm\ boson production in the same phase space is
measured to be \rfid$=(51 \pm 13 \pm 4) \times 10^{-8}$, where the first
uncertainty is statistical, and the second systematic. The
acceptance-corrected observed ratio is \rincl$=(126 \pm 32 \pm 9  {}^{+41}_{-25}) \times
10^{-8}$ where the central value is for the isotropic spin-alignment scenario, and the third
uncertainty represents possible variation due to the unknown spin-alignment.  
The contribution to the total \wpjpsi rate arising from
single parton scattering (double parton scattering subtracted data) is estimated to be \rdpssub$= (78 \pm 32 \pm 22 {}^{+41}_{-25})  \times 10^{-8}$, 
assuming that for DPS the \wpm and prompt \jpsi production factorize completely.
The distribution of the azimuthal angle between the
produced \wpm and \jpsi\ suggests the presence of both SPS and DPS
contributions, and that the DPS estimate accounts for a large fraction of
the observed signal. Comparing the two predictions, the colour singlet
mechanism is expected to be the dominant contribution to the cross section.

Analysis of additional data will likely prove valuable in distinguishing the contributions of colour-singlet and colour-octet predictions to SPS,
and to determine the relative rates of SPS and DPS production. This final state may prove to be a compelling observable for study of
double parton scattering dynamics, as well as for the study of Higgs boson charm couplings
and for ongoing BSM physics search programmes.

\acknowledgments



We thank CERN for the very successful operation of the LHC, as well as the
support staff from our institutions without whom ATLAS could not be
operated efficiently.

We acknowledge the support of ANPCyT, Argentina; YerPhI, Armenia; ARC,
Australia; BMWF and FWF, Austria; ANAS, Azerbaijan; SSTC, Belarus; CNPq and FAPESP,
Brazil; NSERC, NRC and CFI, Canada; CERN; CONICYT, Chile; CAS, MOST and NSFC,
China; COLCIENCIAS, Colombia; MSMT CR, MPO CR and VSC CR, Czech Republic;
DNRF, DNSRC and Lundbeck Foundation, Denmark; EPLANET, ERC and NSRF, European Union;
IN2P3-CNRS, CEA-DSM/IRFU, France; GNSF, Georgia; BMBF, DFG, HGF, MPG and AvH
Foundation, Germany; GSRT and NSRF, Greece; ISF, MINERVA, GIF, DIP and Benoziyo Center,
Israel; INFN, Italy; MEXT and JSPS, Japan; CNRST, Morocco; FOM and NWO,
Netherlands; BRF and RCN, Norway; MNiSW and NCN, Poland; GRICES and FCT, Portugal; MNE/IFA, Romania; MES of Russia and ROSATOM, Russian Federation; JINR; MSTD,
Serbia; MSSR, Slovakia; ARRS and MIZ\v{S}, Slovenia; DST/NRF, South Africa;
MINECO, Spain; SRC and Wallenberg Foundation, Sweden; SER, SNSF and Cantons of
Bern and Geneva, Switzerland; NSC, Taiwan; TAEK, Turkey; STFC, the Royal
Society and Leverhulme Trust, United Kingdom; DOE and NSF, United States of
America.

The crucial computing support from all WLCG partners is acknowledged
gratefully, in particular from CERN and the ATLAS Tier-1 facilities at
TRIUMF (Canada), NDGF (Denmark, Norway, Sweden), CC-IN2P3 (France),
KIT/GridKA (Germany), INFN-CNAF (Italy), NL-T1 (Netherlands), PIC (Spain),
ASGC (Taiwan), RAL (UK) and BNL (USA) and in the Tier-2 facilities
worldwide.

\bibliographystyle{JHEP}

\providecommand{\href}[2]{#2}\begingroup\raggedright\endgroup

\onecolumn
\clearpage 
\begin{flushleft}
{\Large The ATLAS Collaboration}

\bigskip

G.~Aad$^{\rm 48}$,
T.~Abajyan$^{\rm 21}$,
B.~Abbott$^{\rm 112}$,
J.~Abdallah$^{\rm 12}$,
S.~Abdel~Khalek$^{\rm 116}$,
O.~Abdinov$^{\rm 11}$,
R.~Aben$^{\rm 106}$,
B.~Abi$^{\rm 113}$,
M.~Abolins$^{\rm 89}$,
O.S.~AbouZeid$^{\rm 159}$
% ...
\bigskip
\\
% ...
$^{11}$ Institute of Physics, Azerbaijan Academy of Sciences, Baku, Azerbaijan\\
$^{12}$ Institut de F{\'\i}sica d'Altes Energies and Departament de F{\'\i}sica de la Universitat Aut{\`o}noma de Barcelona, Barcelona, Spain\\
% ...
$^{21}$ Physikalisches Institut, University of Bonn, Bonn, Germany\\
% ...
$^{48}$ Fakult{\"a}t f{\"u}r Mathematik und Physik, Albert-Ludwigs-Universit{\"a}t, Freiburg, Germany\\
% ...
$^{89}$ Department of Physics and Astronomy, Michigan State University, East Lansing MI, United States of America\\
% ...
$^{106}$ Nikhef National Institute for Subatomic Physics and University of Amsterdam, Amsterdam, Netherlands\\
% ...
$^{112}$ Homer L. Dodge Department of Physics and Astronomy, University of Oklahoma, Norman OK, United States of America\\
$^{113}$ Department of Physics, Oklahoma State University, Stillwater OK, United States of America\\
% ...
$^{116}$ LAL, Universit{\'e} Paris-Sud and CNRS/IN2P3, Orsay, France\\
% ...
$^{159}$ Department of Physics, University of Toronto, Toronto ON, Canada\\
% ...
$^{s}$ Also at Ochadai Academic Production, Ochanomizu University, Tokyo, Japan\\
% ...
\end{flushleft}


\end{document}